\documentclass[reprint, amsmath,amssymb,aps,letter]{revtex4-1}

\usepackage{float}
\usepackage{graphicx}% Include figure files
\usepackage{dcolumn}% Align table columns on decimal point
\usepackage{bm}% bold math

\usepackage[usenames,dvipsnames]{color}

\usepackage{longtable}
\usepackage{multirow}
\usepackage{comment}

\usepackage{fdsymbol}
\usepackage{manfnt}

\usepackage{soul}
\usepackage[normalem]{ulem}

\begin{document}

\preprint{APS/123-QED}

\title{Influence of Hydration and Dehydration on the Viscoelastic Properties of Snail Mucus by Brillouin Spectroscopy}% Force line breaks with \\

\author{D. F. Hanlon}
\email{dfh031@mun.ca}

\author{M. J. Clouter}
\author{G. Todd Andrews}
\affiliation{Department of Physics and Physical Oceanography, Memorial University of Newfoundland and Labrador, St. John's, NL, Canada, A1B 3X7}

\date{\today}% It is always \today, today,
\begin{abstract}
Brillouin spectroscopy was used to probe the viscoelastic properties of diluted snail mucus at GHz frequencies over the range -11 $^\circ$C $\leq T \leq$ 52 $^\circ$C and of dehydrated mucus as a function of time. Two peaks were observed in the spectra for diluted mucus: the longitudinal acoustic mode of the liquid mucus peak varies with dilution but fluctuates around the typical value of 8.0 GHz. A second peak due to ice remained unchanged with varying dilution and was seen at 18.0 GHz and appeared below the dilutions ``freezing" point depression. Only a single peak was found in all the dehydrated mucus spectra and was also attributed to the longitudinal acoustic mode of liquid mucus. Anomalous changes in the protein concentration dependence of the frequency shift, linewidth, and ``freezing" point depression and consequently, hypersound velocity, compressibility, and apparent viscosity suggest that the viscoelastic properties of this system is influenced by the presence of water. Furthermore, this research uncovered three unique transitions within the molecular structure. These transitions included the first stage of glycoprotein cross-linking, followed by the steady depletion of free water in the system, and eventually resulted in the creation of a gel-like state when all remaining free water was evaporated.

\begin{description}
\item[PACS numbers]
May be entered using the \verb+\pacs{#1}+ command.
 
\end{description}
\end{abstract}

\pacs{Valid PACS appear here}
\maketitle
\section{Introduction}

In recent years, there has been a growing interest in exploring the mechanical properties of polymer-water solutions, driven by their significance in understanding the functional characteristics of proteins \cite{bailey2019brillouin, comez2012, comez2016, lupi2011}. Previous research however has primarily focused on rheological studies of natural snail mucus \cite{denn1980,denn1984}. Recent studies on natural snail mucus used Brillouin light scattering spectroscopy to investigate the effect of temperature on its viscoelastic properties \cite{hanlon2023temperature}. The findings in that work revealed that snail mucus exhibits increased viscoelastic properties, such as the sound velocity, storage and loss moduli, as well as apparent viscosity, in comparison to water. The study also discovered a phase transition in which the viscous liquid state of snail mucus shifts to a condition in which both liquid mucus and solid ice phases coexist. This work sheds light on the peculiar behaviour of snail mucus under different temperature conditions. The results of this study imply that the glycoprotein-water interaction is responsible for the increased viscoelastic behaviour and phase change observed in snail mucus. The results of this Brillouin scattering investigation, as well as prior rheological experiments, clearly indicate that the cross-linked network of glycoproteins in snail mucus is the major reason of the greater mechanical properties reported in snail mucus \cite{denn1980,denn1984,hanlon2023temperature}. The previous Brillouin light scattering study, while it does investigate the temperature dependence of the system, it does not explore how the glycoprotein concentration influences the mechanical properties of snail mucus. While a number of studies in the past have explored the viscoelastic properties of polymer-water systems, the impact of glycoprotein concentration on the viscoelastic properties such as complex modulus, sound velocity, and apparent viscosity still remains poorly understood \cite{bailey2019brillouin,comez2012,comez2016,lupi2011}.

Snail mucus is a polymer-water solution composed of long chains of high molecular weight glycoproteins in water. These glycoproteins are known to to cross-link in the mucus. Protein concentrations in these solutions were previously reported in the literature to range from 3 - 7\% proteins by weight, with the remaining composition consisting of water \cite{denn1980,denn1984,verd1987}. The interaction between proteins and water plays a significant role in the functionality of these polymer-water solutions and holds important implications for various technological advancements. Active research is being conducted in the field of hydrogels, which are being explored for their potential as non-toxic medical adhesives \cite{li2017}. A comprehensive understanding of the limitations associated with these hydrogels is essential for their effective fabrication in medical applications. Furthermore, the cryoprotectants field is actively exploring polymer-water solutions, particularly antifreeze glycoproteins, as potential substitutes for traditional cryoprotectants. These solutions exhibit the capability to significantly reduce the freezing temperature, even at low concentrations, thereby presenting exciting prospects for cryopreservation applications \cite{bouv2003,kwan2020}. In recent years, there has been a increased research focused on protein-water solutions to gain a deeper understanding of the dynamics of water in the vicinity of proteins with emphasis on understanding the interesting behavior of water molecules surrounding proteins and their interactions \cite{lupi2011,comez2012,comez2016,monaco2001glass,ebbi2010}.

In this paper, Brillouin light scattering spectroscopy was used to characterize the temperature dependence of diluted snail mucus as function of glycoprotein concentration over the range  -11 $^\circ$C to 52 $^\circ$C. We report values for sound velocity, storage (bulk) moduli, apparent viscosity, and hypersound absorption for all dilutions as a function of temperature. Additionally, we show the influence of hydration on the previously discovered freezing point depression attributed by the presence of glycoproteins \cite{hanlon2023temperature}. Furthermore, we investigate the influence dehydration has on the viscoelastic properties of snail mucus by allowing the mucus to naturally dehydrate over the course of $\sim$ 400 hours. The results of these two experiments build upon previously published data on the viscoelastic properties  of natural snail mucus, offering new insights into the physics of aqueous glycoprotein solutions and shedding light on the role of glycoproteins in the phase behavior of these systems \cite{hanlon2023temperature}. In a wider context, the present study enhances our understanding of phonon dynamics in complex fluids, unraveling the complex interplay between water and biomacromolecules, and illuminating the significance of glycoproteins in biological systems. 

\section{Methodology}
\subsection{Brillouin Light Scattering Spectroscopy}
\subsubsection{Brillouin Scattering in Liquids}

% \hl{Throughout paper important to keep notation consistent with first mucus paper, particularly for your thesis reviewers.  Use here same symbols used for quantities that appear in first paper (and maintain this consistency in notation for subsequent papers that end up in thesis.} 

Brillouin light scattering spectroscopy is an inelastic light scattering technique which measures random thermal density fluctuations in liquids.  The intensity of light inelastically scattered by aqueous solutions given by I$_q$($\omega$), with exchanged wave vector $q$, is dominated by density fluctuations in the medium, which can be derived from the theoretical framework of hydrodynamics \cite{comez2012progress, montrose1968brillouin,tao1992brillouin}. The intensity can be represented with the following expression,

\begin{equation}
    I_q (\omega) \propto S_q  = \frac{S_q M_0}{\pi \omega_B} \frac{M^{\prime\prime}}{[M^{\prime} - \rho \omega_B^2 / q^2]^2 + [M^{\prime\prime}]^2},
\end{equation} 
where $S_q$ is the static structure factor, $\rho$ is the mass density, $M_0$ is the relaxed (zero-frequency) longitudinal acoustic modulus, $q$ = (4$\pi/\lambda$)$n \sin(\theta/2)$, $n$ is the refractive index of the medium, (here assumed constant at $n = 1.34$ \cite{gugl2021}) and $\theta$ is the scattering angle. Additionally, $\omega$ is the angular frequency of the density fluctuations, and $\omega_B$ is the angular frequency of the density fluctuations obtained through Brillouin scattering. The complex longitudinal modulus, $M(\omega) = M^{\prime}(\omega) + i M^{\prime\prime}(\omega)$, often referred to as the complex modulus, can be obtained directly via Brillouin spectroscopy using the following relations \cite{bailey2019brillouin},
 \begin{equation}
    M = M^{\prime} + iM^{\prime\prime} = \frac{\lambda_i^2 \rho}{4 n^2} f_B^2 + i\frac{\lambda_i^2 \rho}{8 \pi n^2} f_B \Gamma_B,
    \label{eq:longmod}
\end{equation}
where $\Gamma_B$ is the full width at half maximum (FWHM) and $M^{\prime}$ is referred as storage modulus and is a measure of how much energy is stored elastically in the system. The loss modulus, $M^{\prime\prime}$, is a measure of how much energy is lost through heat in the system. $M^{\prime}$ and $M^{\prime\prime}$ can be expressed in terms of sound velocity $v$ and apparent viscosity $\eta = 4\eta_{s}/3 + \eta_{b}$, where $\eta_{s}$ and $\eta_{b}$ are the shear and bulk viscosity, respectively, can be represented by the following,
\begin{equation}
    M^{\prime} = \rho v^2
\end{equation}
and 
\begin{equation}
    M^{\prime\prime} = 2\pi \eta f_B
\end{equation}
\noindent
For a $180^{\circ}$ backscattering geometry used in the present work, application of energy and momentum conservation to the scattering process reveals that the phonon velocity, $v$, and frequency shift of the incident light, $f$, are related by  
\begin{equation}
v = \frac{f\lambda_i}{2n},
\label{eqn:brillouineqn}
\end{equation}
where $n$ is the refractive index of the target material at the incident light wavelength $\lambda_{i}$. 

The spectral width of the Brillouin peaks is determined by the time that each fluctuation interacts with the incident light, thus measuring its lifetime (thermal relaxation) or the attenuation of these fluctuations \cite{dil1982}. Using both the frequency shift and FWHM, the frequency-independent sound absorption coefficient can be calculated using the following \cite{rouc1976},
\begin{equation}
\frac{\alpha}{f^2} = \frac{\Gamma_B}{2vf_B^2}.
\label{eq:absorp}
\end{equation}
where, $f_B$, and $\Gamma_B$ are directly obtained from the Brillouin spectra, and $v$ is is calculated from Eq. \ref{eqn:brillouineqn}.

\subsubsection{Apparatus}
Brillouin spectra were obtained using a $180^\circ$ backscattering geometry using the experimental setup described previously in Ref. \cite{andr2018}. A Nd:YVO$_4$ single mode laser provided light of wavelength $\lambda_{i} = 532$ nm and power of 100 mW, which was directed onto the samples inside a temperature-controlled sample chamber described in great detail previously \cite{hanlon2023temperature}. A high-quality anti-reflection-coated camera lens of focal length $f=5$ cm and $f/\# = 2.8$ served to both focus incident light onto the sample and to collect light scattered by it.  After exiting this lens, the scattered light was focused by a 40 cm lens onto the 450 $\mu$m-diameter input pinhole of an actively-stabilized 3+3 pass tandem Fabry-Perot interferometer (JRS Scientific Instruments). The interferometer had a free spectral range  of 30 GHz, and possessed a finesse of approximately 100. The light transmitted by the interferometer was directed onto a pinhole with a diameter of 700 $\mu$m and subsequently detected by a low-dark count ($\lesssim 1$ s$^{-1}$) photomultiplier tube where it was converted to an electrical signal and sent to a computer for storage and display. 

\subsubsection{Spectral Peak Analysis}
Peak parameters including frequency shifts, and linewidths (FWHM), were obtained by fitting Lorentzian functions to the Stokes and anti-Stokes peaks and averaging the resulting parameters of best-fit. To obtain spectral linewidths, the instrumental linewidth of 0.3 GHz was subtracted from FWHM values obtained from the fits. Estimated uncertainties in peak parameters were  also obtained from the uncertainty in the fits. 

\subsection{Attenuated Total Reflectance IR Spectroscopy} 
\subsubsection{Theory}

Fourier Transform Infrared (FTIR) spectroscopy, a type of vibrational spectroscopy, is a technique used to obtain an infrared spectrum of absorbance or emission of a solid, liquid or gas \cite{griffiths1983fourier}. This technique allows for the identification and characterization of various chemical compounds which makes it useful in understanding the structure of complex liquid systems and is often used in conjunction with other spectroscopy techniques in order to characterize different samples \cite{arunkumar2019ftir,lewis2013prediction,spiridon2011structural,griffiths1983fourier}. The technique employs a Michelson interferometer to acquire IR spectra of the material. FTIR data is commonly gathered through an attenuated total reflectance (ATR) method. The ATR technique utilizes a crystal prism, such as diamond or zinc selenide, in direct contact with the sample \cite{grdadolnik2002atr}. In this method, the infrared radiation is directed onto the prism at a precise angle, subsequently undergoing multiple internal reflections by the prism before being ultimately absorbed by the sample \cite{grdadolnik2002atr}. 

% Compared to traditional transmission IR spectroscopy, this technique yields a significantly improved signal-to-noise ratio. \hl{Is it true that all of the IR is absorbed by the sample?}

\subsubsection{Apparatus}
All data was collected using a Vertex 70v vacuum Fourier transform infrared (FTIR) spectrometer (Bruker, Billerica, MA, USA) with Platinum Diamond attenuated total reflectance (ATR) attachment and a globar (blackbody) light source. Data was collected using a KBr beam splitter (4000–400 cm$^{-1}$) with 2 cm$^{-1}$ resolution. 

Before each measurement, a background spectra was first collected to calibrate the system. Each sample was collected by using a pipette to extract the mucus, which was then directly placed onto the ATR crystal. Approximately 0.5 ml of mucus was used for each sample to ensure complete coverage of the crystal surface. 

% \hl{How was sample put onto apparatus for measurement?  Approximate quantity of sample? + other such details, if any.}

\subsubsection{Spectral Analysis}
\begin{figure}[t]
\centering
\includegraphics[width=1\linewidth]{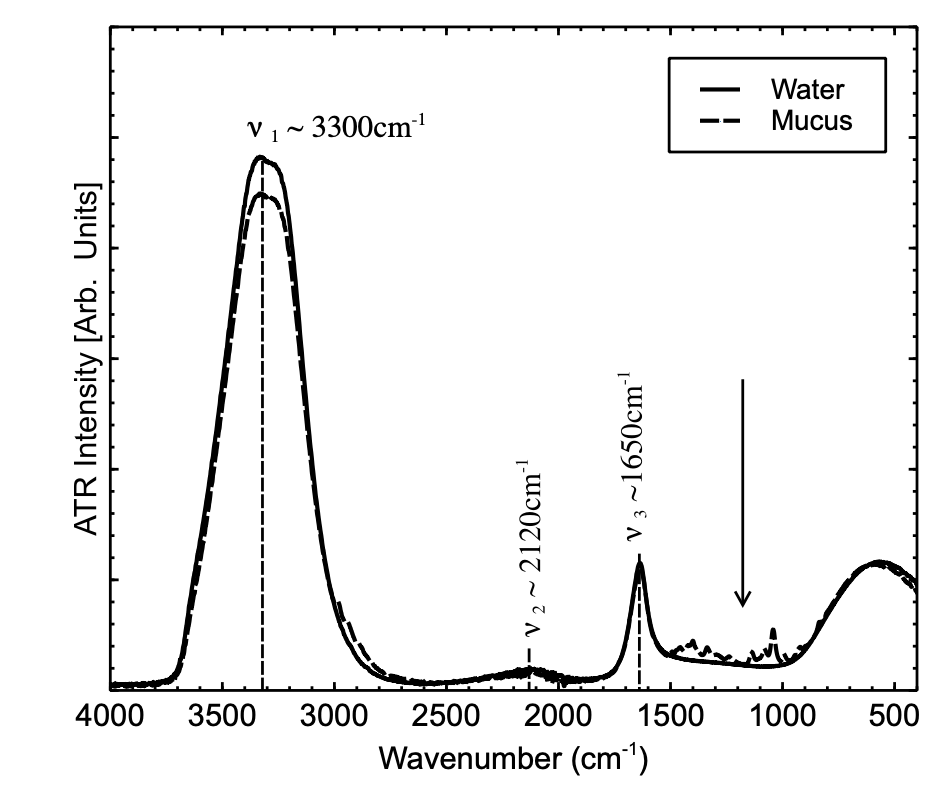}
\caption{ATR spectra collected on ultra-pure water and pure snail mucus at room temperature.}
\label{fig:FTIR_Mucus_Water}
\end{figure}

The data collection procedure involved initially collecting FTIR spectra on ultra-pure water, as depicted in Fig. \ref{fig:FTIR_Mucus_Water}. This spectra served as a ``baseline" for all subsequent spectra obtained for diluted and dehydrated mucus. ATR spectra initially collected on pure snail mucus and showed very good agreement with spectra previously obtained for gastropod mucus. Furthermore, ATR spectra were collected on all diluted mucus samples. As for the dehydrated mucus, spectra were only collected after a noticeable change in Brillouin data was observed. Additionally, the presence of proteins in the snail mucus is indicated by the arrow in Fig. \ref{fig:FTIR_Mucus_Water}.

FTIR data in this work are displayed according to the common convention of ATR intensity versus wave number, where the wave numbers are displayed with decreasing values from left to right. The concentration of pure mucus, diluted and dehydrated mucus was estimated by comparing the integrated intensity of water absorbance bands to the same bands visible in spectra of diluted and dehydrated mucus. Figure \ref{fig:FTIR_Mucus_Water} shows FTIR spectra of ultra-pure water and pure mucus. Both spectra show three prominent absorbance bands $\nu_1$, $\nu_2$ and $\nu_3$ (see Fig. \ref{fig:FTIR_Mucus_Water} due to the symmetric O-H stretching, O-H bending (scissors) vibration and a combination band of $\nu_2$ plus an additional motion of molecule. These absorbance bands act as a fingerprint for water and can help in identifying the presence of water in various systems. Additionally, we see the presence of absorbance bands likely due to the proteins, as indicated by the arrow in Fig. \ref{fig:FTIR_Mucus_Water}.

\subsubsection{Concentration Determination}

\begin{figure}[t]
\centering
\includegraphics[width=1\linewidth]{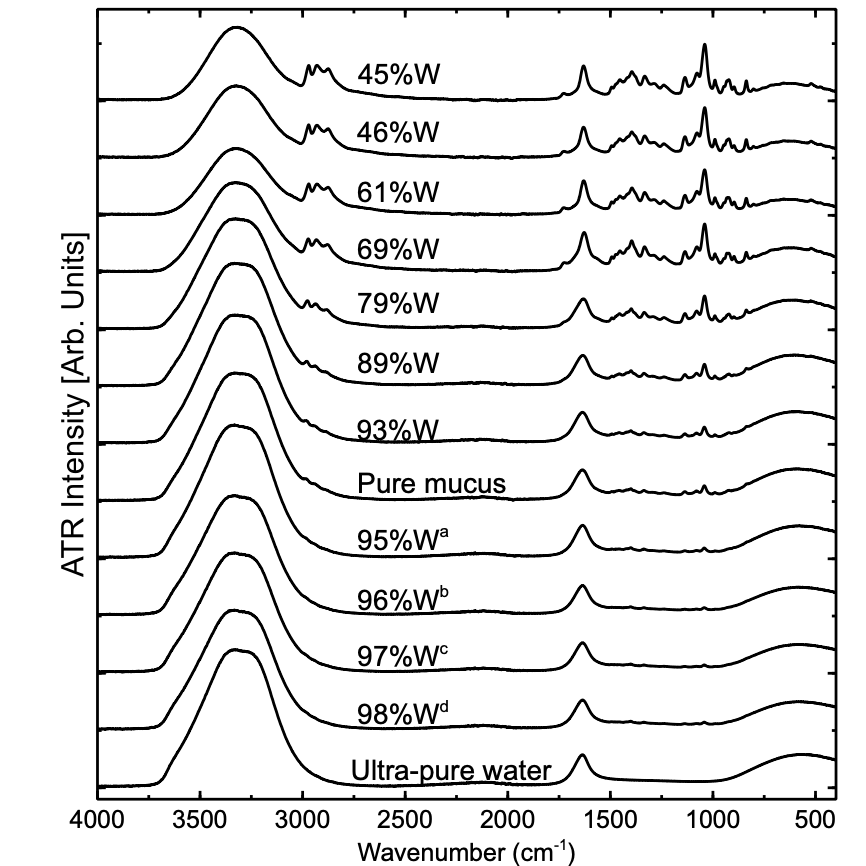}
\caption{ATR-FTIR spectra of natural snail mucus. ATR spectra collected on diluted snail mucus are indicated by the superscripts $a-d$. The remaining spectra were collected on dehydrating mucus as a function of time.}
\label{fig:FTIR_Time}
\end{figure}

One of the primary objectives of utilizing FTIR in this study was to estimate the concentration of water and proteins within the snail mucus. Prior research suggests that an accurate concentration can be determined by comparing the integrated intensity of the spectral water peak with that of an unidentified peak of some system containing water \cite{venyaminov1997water}. For example, Fig. \ref{fig:FTIR_Mucus_Water} shows very similar ATR spectra, and by comparing the absorbance bands of  water, with those peaks at the same position in the mucus we can determine the concentration of protein and water present. The intensity of these absorbance peaks gives a sense of the abundance of molecules absorbing the IR radiation \cite{larkin2017infrared}. To obtain the integrated intensity, a Gaussian profile was used to fit to the absorbance bands shown in Fig. \ref{fig:FTIR_Time} due to its ability to properly fit the bands of cross-linked networks \cite{bradley2007curve} (see Table S1 in Supplementary file for values extracted from ATR spectra).  

In general, the shifts of all vibrational modes remained relatively constant. The FWHM for the $\nu_1$ generally decreases with decreasing concentration, $\nu_2$ remains constant for most of the concentrations, and $\nu_3$ increases with increasing concentration for those spectra where the peak is observed. The most obvious changes in Fig. \ref{fig:FTIR_Time} is with the spectral intensity. The vibrational mode $\nu_1$ is the most intense throughout all spectra, and shows a general decrease from ultra-pure water to the spectra containing 45\% water. The intensity of the $\nu_2$ vibrational mode is relatively constant, showing only slight variations between 15-40 counts over the whole range studied. As well, the intensity of the vibrational mode at 2130 cm$^{-1}$ remains constant throughout. 

\subsubsection{Density Determination}
The density of each diluted and dehydrated mucus solution was determined using water and protein concentrations determined from ATR analysis and the following expression for the density of a two-component solution \cite{bailey2019brillouin}:

\begin{equation}
    \rho = \frac{m_W + m_M}{\frac{m_w}{\rho_w} + \frac{m_M}{\rho_M}},
    \label{eq:density}
\end{equation}
\noindent
where $m_W$ and $m_M$ denote the mass of water and dry mucus, respectively, and $\rho_W = 0.997$ g/cm$^{3}$ \cite{ocon1967} and $\rho_M = 1.571$ g/cm$^{3}$ are the corresponding mass densities.  The dry mucus density $\rho_M$ was obtained from Eq. \ref{eq:density} using the known density of natural snail mucus ($\rho =1.040$ g/cm$^3$) determined in a previous study \cite{hanlon2023temperature} and compositional information obtained through ATR analysis.

%To calculate densities for various dilutions and dehydrations, Equation \ref{eq:density} was employed, utilizing the aforementioned mass densities and the water and protein concentration determined from ATR analysis. This enabled the determination of densities for all dilutions and dehydrations.

% \begin{table*}[ht]
% \centering
% \caption{Sample Analysis}
% \begin{tabular}{cccccc}
% \hline
% Sample & $\nu_1$ & $\nu_2$ & $\nu_3$ &FWHM & Intensity \\
% \hline
% Sample 1 & 1500 & 0.5 & 10 \\
% Sample 2 & 1420 & 0.4 & 15 \\
% Sample 3 & 1550 & 0.6 & 12 \\
% Sample 4 & 1480 & 0.3 & 8 \\
% Sample 5 & 1600 & 0.8 & 9 \\
% Sample 6 & 1370 & 0.7 & 11 \\
% \hline
% \end{tabular}
% \end{table*}

% Furthermore, the distinction between ATR data for ultra-pure water and snail mucus lies in the existence of additional absorbance bands found within the region 1700 cm$^{-1}$ to 1000 cm$^{-1}$. This particular region enables us to estimate the amino acid types by comparing the vibrational modes with previously documented vibrations in the literature pertaining to proteins (cite papers here). Table XX below gives the peak positions for absorbance bands found in snail mucus along with the likely assignment for each band.

% \begin{table}[b]
%   \centering
%   \caption{Overview of ATR-FTIR absorbance bands found in a characteristic H$_2$O sample}
%   \begin{tabular}{lll}
%     \hline\hline
%     Assignment & Band Position [cm$^{-1}$]& Remarks \\
%     \midrule
%     Amide I & 1650-1700 & Mainly C=O stretching \\
%     Amide II & 1550-1600 & N-H bending and C-N stretching \\
%     Amide III & 1220-1300 & C-N stretching and N-H bending \\
%     \hline\hline
%     \label{tab:OverviewWATER-FTIR}
%   \end{tabular}
% \end{table}

\section{Dilution Experiments} 
\subsection{Sample Preparation} 
The diluted snail mucus samples utilized in this study were prepared by combining natural snail mucus with ultra-pure water. To prevent dehydration, the mucus was initially stored in a sealed container. Using a syringe, 1 ml of mucus was extracted from the parent container and transferred to a glass sample cell that was sealed at one end. Ultra-pure water was added to the cell in 0.5 ml increments for each dilution. After water was added, the mucus-water solution was stirred with a stirring stick for 10 minutes to ensure proper mixing. This standardized procedure was followed for all dilutions to maintain consistency.

\begin{table*}[t]
  \centering
  \caption{Sample descriptions for Diluted Snail Mucus}
  \label{tab:systems}
  \setlength{\tabcolsep}{14pt} % Increase the spacing between columns
  \begin{tabular}{ccccc}
    \hline \hline
    &Dilution \#1 & Dilution \#2 & Dilution \#2 & Dilution \#4 \\
    \hline
   Total Volume (mL) & 1.5 $\pm$ 0.1 & 2.0 $\pm$ 0.1& 2.5 $\pm$ 0.1 & 1.5 $\pm$ 0.1 \\
    Protein Concentration (wt \%) &5 $\pm$ 0.7 & 4 $\pm$ 0.6 & 3 $\pm$ 0.5 & 2 $\pm$ 0.2 \\
    Water Concentration (wt \%) & 95 $\pm$ 0.7 & 96 $\pm$ 0.6 & 97 $\pm$ 0.5 & 98 $\pm$ 0.2 \\
    Density (g/cm$^3$) & 1.027 $\pm$ 0.007 &  1.020 $\pm$ 0.005 &  1.017 $\pm$ 0.005 &  1.013 $\pm$ 0.003 \\
    \hline\hline
  \end{tabular}
\end{table*}

Following the mixing of solutions, the samples were left undisturbed to equilibrate for 6 hours. Brillouin spectra were collected as a function of temperature after each dilution. It should be noted that the glass sample cell had physical limitations, and the maximum amount of water that could be added was 2.5 ml. Additionally, an alternative sample was prepared by initially adding 0.5 ml of mucus first and then adding 1.5 ml of water. Once each sample was prepared, the glass sample cell was placed into the previously discussed temperature-controlled sample chamber \cite{hanlon2023temperature}.

\subsection{Results: Raw Spectral Signatures}
% \hl{In R\&D sections (both here and for Dehydration Experiments) it is usually, but not always, a good idea to  put figures (and tables if appropriate) with raw data along with associated text sections first and follow these with figures/tables/text that include interpretation/analysis of raw data.  Just a reminder in case you have not considered this organizational point.}

\begin{figure}[t]
\centering
\includegraphics[width = 1\linewidth]{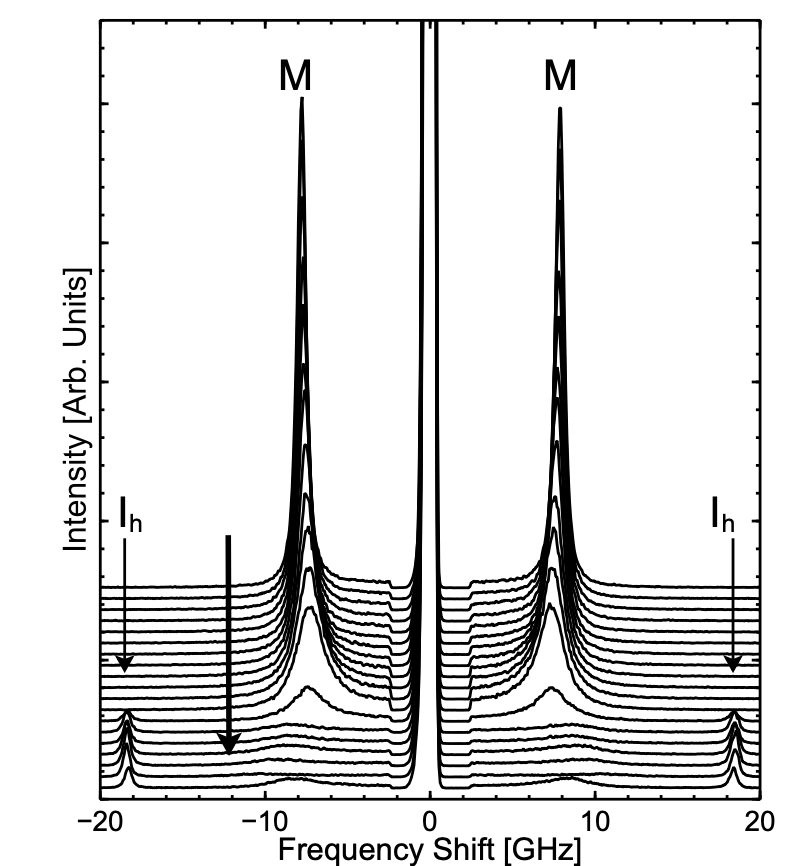}
\caption{Temperature dependence of Brillouin spectra for diluted gastropod mucus with 5 wt\% protein and 95 wt\% water (Dilution \#1 in Table \ref{tab:systems}). Direction of decreasing temperature is indicated by the large arrow second from left. \textbf{M} and \textbf{I$_{\mathrm{h}}$} identify peaks due to mucus and ice, respectively. }
\label{fig:BrillouinSpectra_Dilutions}
\end{figure}

\begin{figure}[t]
\centering
\includegraphics[width=1\linewidth]{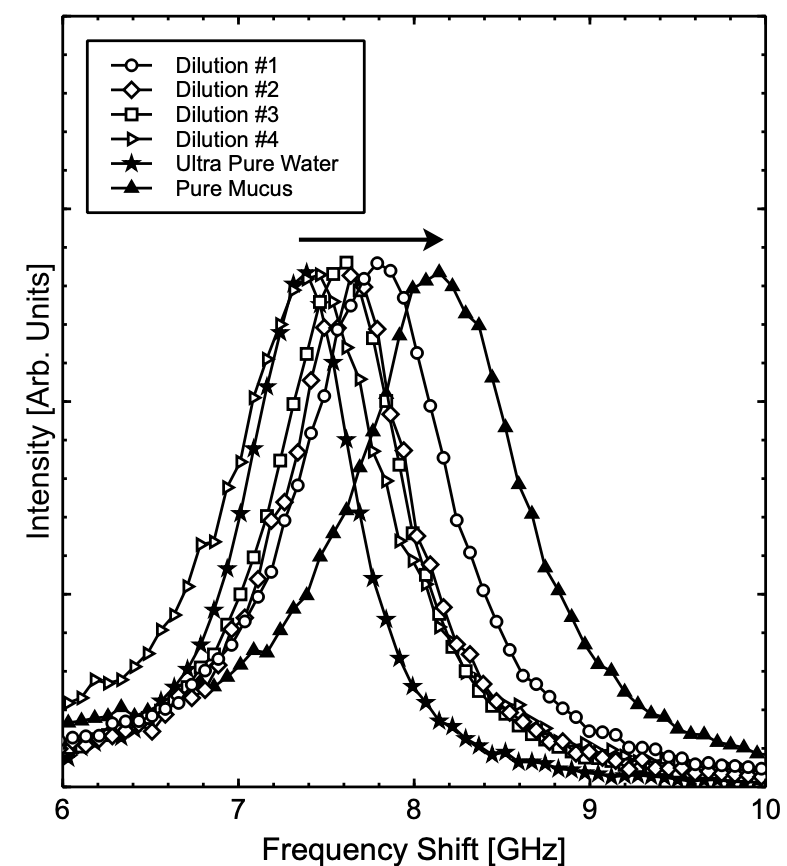}
\caption{Room temperature Brillouin spectra (Anti-Stokes) of all diluted gastropod mucus samples. Arrow represents direction of increasing protein concentration.}
\label{fig:BrillouinSpectra_AStokes}
\end{figure}
Figure \ref{fig:BrillouinSpectra_Dilutions} shows representative series of Brillouin spectra  collected on diluted snail mucus for temperature ranging from -11.0$^\circ$C - 50.0$^\circ$C. Figure \ref{fig:BrillouinSpectra_AStokes} shows the anti-stokes Brillouin peak for liquid mucus for all dilutions at room temperature displaying the general difference in spectra for each dilution. As evident from Fig. \ref{fig:BrillouinSpectra_AStokes}, as the dilution increases, the Brillouin shift decreases.
\begin{figure}[t]
\centering
\includegraphics[width=1\linewidth]{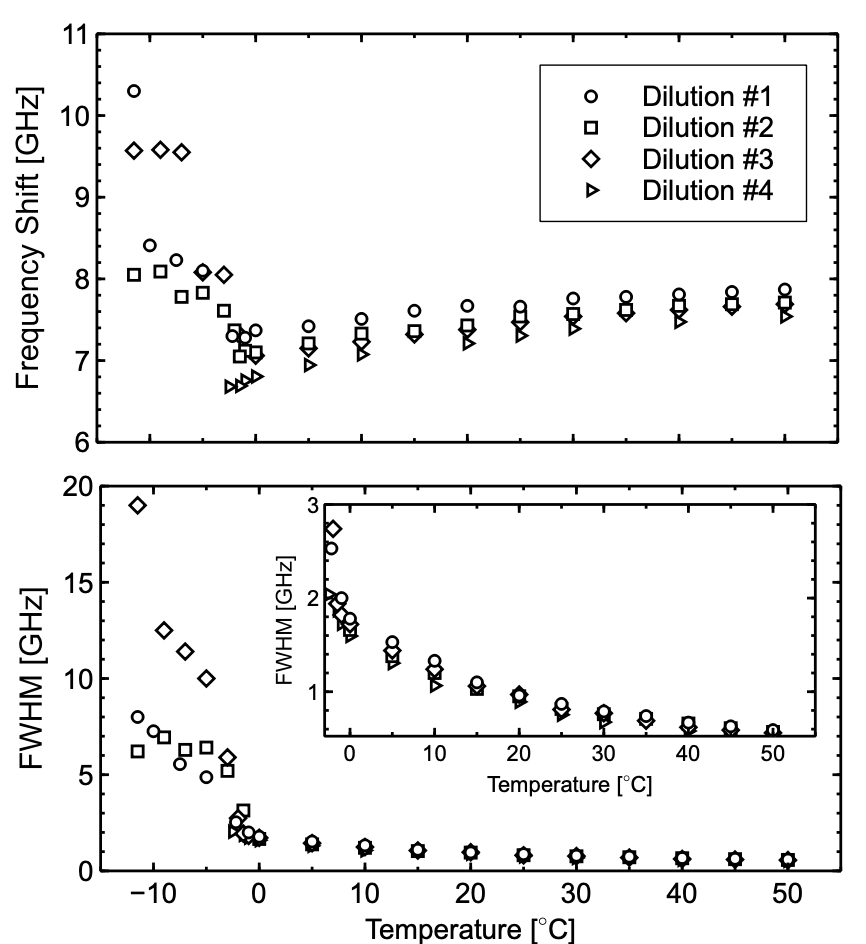}
\caption{Plot of frequency shift, and FWHM of diluted snail mucus samples as a function of temperature.}
\label{fig:FrequencyShift_FWHM_Dilutions}
\end{figure}
In all dilutions there were two sets of Brillouin peaks present in the spectra, one of which occured at a shift of $\sim$ 8.0 GHz which decreased to a value similar to that of water as the dilution increased. This peak was present over the entire temperature range for all dilutions, however the spectral intensity of this peak decreased with increasing dilution for temperatures below the corresponding ``freezing" point.  A second peak occured at approximately 18.0 GHz and has previously been deemed due to the longitudinal acoustic modes of ice I$_h$ \cite{hanlon2023temperature}. Throughout this manuscript, the term ``freezing" point is to indicate the point at which this ``ice peak" appears in the spectrum.

% shows the frequency shift of Brillouin peaks for all dilutions attributed to the longitudinal acoustic mode of liquid gastropod mucus previously reported \cite{hanlon2023temperature}.

Figure \ref{fig:FrequencyShift_FWHM_Dilutions} shows the frequency shift as a function of temperature for all dilutions.  For all dilutions there is a slight increase in the frequency shift for increasing temperatures greater than its freezing point. Below their respective freezing points however, dilutions 1-3 (see Table \ref{tab:systems}) all show a rapid increase in frequency shift with decreasing temperature. Dilution 4 on the other hand shows that below its "freezing" point, the frequency shift follows a similar trend to that of pure water, and decreases with decreasing temperature. This result suggests that there is a transition occurring in dilution\#4 that is disrupting the network of glycoproteins and water molecules present in snail mucus. This is further supported by the difference in frequency shift between the most dilute (Dilution \#4) and least dilute (Dilution \#1). This shows that for higher protein concentration (Dilution \#1), the frequency shift resembles that of pure mucus Ref.~\cite{hanlon2023temperature}, while the lowest protein concentration (Dilution \#4) resembles that of ultra pure water Ref.~\cite{hanlon2023temperature}. Previous concentration dependent Brillouin studies on polymer gels shows a similar trend, that is at very low polymer concentrations the frequency shift is comparable to pure water, and at high concentrations the frequency shift is drastically different \cite{zhao1997brillouin,scheyer1997relaxations, pochylski2006structuring}. 

% This is likely due to this dilution having the greatest water content out of all dilutions. A similar phenomenon has been seen in very dilute polymer-water solutions previously \cite{conde1982analysis}.

Figure \ref{fig:FrequencyShift_FWHM_Dilutions} shows the FWHM as a function of temperature for all dilutions which for T $\geq$ 0 $^\circ$C decreases with increasing temperature. Below, this temperature the FWHM in general increases. The inset figure shown in Fig. \ref{fig:FrequencyShift_FWHM_Dilutions} shows the FWHM for temperature region T $\geq$ 0 $^\circ$C. In this temperature region, the FWHM for all dilutions shows slight variations between one another which become more apparent towards T = 0 $^\circ$C, and practically identical near T = 50.0 $^\circ$C. In general, the FWHM decreases with increasing dilutions, this is best illustrated by the series of values at T = 5.0$^\circ$C. This is consistent with the observation in the frequency shift data, that is as the dilution increases, the FWHM approaches values close to ultra pure water. Previous Brillouin scattering studies on aqueous polymer solutions had shown that polymer concentration has a significant influence on the FWHM \cite{bailey2019brillouin, van2000structural, pochylski2006structuring}. In fact, Ref. \cite{bailey2019brillouin} had stated that the polymer concentration has the largest influence on the Brillouin linewidth. Although the concentrations for this dilutions experiments are similar, there is a noticeable change in the FWHM between each dilution.

Following from the Brillouin frequency shift and linewidth, values of hypersound velocity, viscosity and sound absorption can be directly calculated using the raw spectral signatures obtained from Brillouin spectra.

\subsection{Results: Viscoelastic Behaviour}
\subsubsection{Sound Velocity}

\begin{figure}[t]
\centering
\includegraphics[width = 1\linewidth]{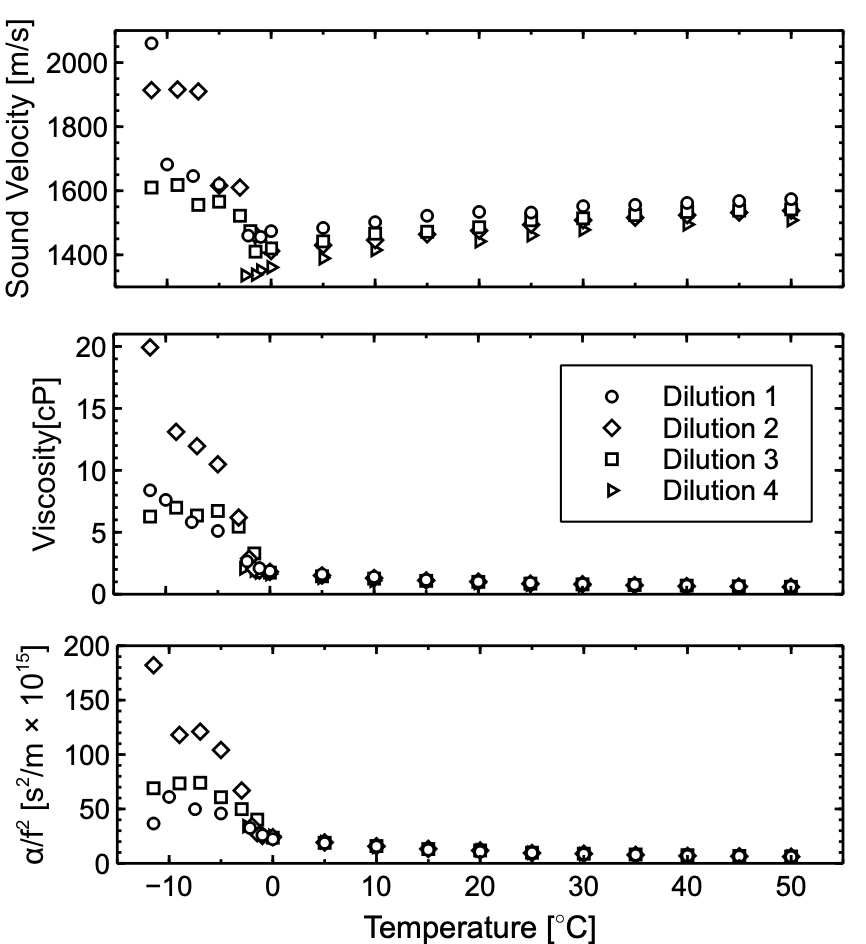}
\caption{Hypersound velocity, apparent viscosity, and sound absorption for diluted mucus as a function of temperature.}
\label{fig:Mucus_Dilution_Comp_Visco}
\end{figure}

Figure \ref{fig:Mucus_Dilution_Comp_Visco} shows the sound velocity for diluted mucus as a function of temperature and was calculated using Eq.. \ref{eqn:brillouineqn}. From examining the hypersound velocity obtained by Brillouin scattering can provide valuable insights into the viscoelastic properties within the diluted snail mucus samples. In general, the sound velocity exhibits interesting variations with temperature across the dilutions. At temperatures above the freezing point of each dilution, there is a gradual increase in hypersound velocity as the temperature increases. This behavior suggests a typical thermal expansion, where the molecules within the mucus exhibit increased vibrational energy and contribute to the propagation of sound waves at higher velocities. However, below their respective freezing points, dilutions 1-3 illustrate a pronounced increase in sound velocity with decreasing temperature. This notable change implies a structural transformation in the snail mucus network, likely induced by addition of ice in mucus. The transition occurring in these dilutions affects the arrangement and interaction of glycoproteins and water molecules, leading to an alteration in the sound velocity. In comparison to the other dilutions, dilution 4 demonstrates a distinct variation in sound velocity at temperatures below its freezing point. The sound velocity in dilution 4 decreases with decreasing temperature, similar to how it does in pure water. This finding implies that a phase transition is observed in dilution 4 which indicate a break in the network of glycoproteins and water molecules in snail mucus.

% The trends observed in this data are analogous to those observed in the frequency shift versus temperature plot. That is, as dilution increases, the sound velocity decreases. Likewise, for Dilutions 1-3, the trend observed below the freezing point, is essentially the same. Dilution 4, however shows an overall decrease in the sound velocity with decreasing temperature, similar to what has been observed for water previously \cite{hanlon2023temperature}.

\subsubsection{Apparent Viscosity}
\begin{figure}[t]
\centering
\includegraphics[width=1\linewidth]{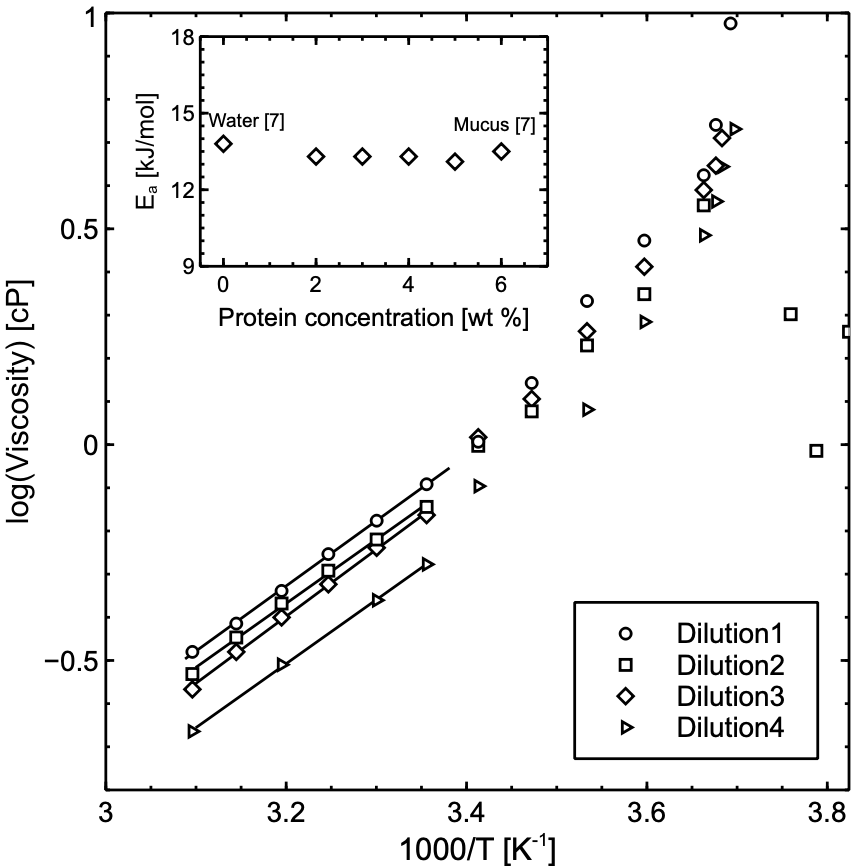}
\caption{Natural logarithm of apparent viscosity of diluted snail mucus samples versus temperature. Solid lines: Best fits of Eq.~\ref{eq:lneta} to high temperature data. Inset: Activation energy $E_a$ obtained from the linear fits as a function of protein concentration.}
\label{fig:lneta}
\end{figure}

Figure \ref{fig:Mucus_Dilution_Comp_Visco} shows the apparent viscosity for the diluted mucus sample as a function of temperature. In general, for T $\geq$ 0$^\circ$C the apparent viscosity decreases with increasing temperature for all dilutions. Although it is hard to tell from Fig. \ref{fig:Mucus_Dilution_Comp_Visco}, there is a systematic decrease in viscosity with increasing dilution. Below T = 0$^\circ$C, the apparent viscosity does not show any consistent behaviour, likely due to the presence of ice crystallites in this region.

Fig. \ref{fig:lneta} shows the natural logarithm of the apparent viscosity of diluted snail mucus as a function of temperature calculated directly from Eq.. \ref{eq:longmod} using the density (see Table \ref{tab:systems}) and FWHM. The viscosity decreases with increasing temperature in a similar manner to what has been previously reported for snail mucus along with other polymer-water systems \cite{so1994,so1995,sirv1993,sirv1994}. 
% The apparent viscosity of each dilution was calculated using Eq. \ref{eq:longmod} along with values of $f_B$ and FWHM. Figure \ref{fig:lneta} shows the batural logarithm of the apparent viscosity for dilutions as a function of 1000/T. the  the activation energy E$_a$ was extracted by performing a linear fit for the T $\geq$ 22$^\circ$C which has previously shown the capability to provide key information into the structural and density relaxation times for these protein -water solutions. 

In the high temperature region (25$^\circ$C $\leq$ T $\leq$ 50$^\circ$C), $\ln (\eta)$ depends linearly on $1/T$ so we fit an Arrhenius relationship of the form

\begin{equation}
\eta = \eta_0 e^{E_a/k_B T} 
\label{eq:lneta}
\end{equation}
to this data and extracted the activation (enthalpy) energy $E_a$.  Here, $\eta_0$ is a prefactor that contains the entropic contribution to the viscosity and $k_B$ is the Boltzmann constant \cite{lupi2011}.  The resulting values for the activation energy $E_a$ of all diluted mucus have values of $\sim$ 13.3 $\pm$ 0.3 kJ/mol (see Table \ref{tab:ViscoFit}). Considering the data shown in Fig \ref{fig:lneta}, dilutions 1 - 3 all exhibit similar values of the natural logarithm of apparent viscosity as the temperature decreases. However, the value for dilution 4 is approximately 2\% smaller than the other dilutions and is very similar to the value of water previously reported \cite{hanlon2023temperature}. This observed deviation we see in the natural logarithm of apparent viscosity  between dilutions 3 and 4 is an indication of a structural change occurring in the mucus at the low protein concentration.

Because the activation energy E$_a$ is consistent across all dilutions, as well as prior findings for snail mucus, the change in apparent viscosity was attributed to the pre-entropic component $\eta_0$ \cite{hanlon2023temperature}. However, that study did not investigate the influence of protein concentration on the apparent viscosity or the activation energy. The concentration dependence of activation energy for polymer-water solutions previously studied displayed the same findings presented here \cite{lupi2011,comez2016,monaco2001glass,comez2012}. That is, as concentration increases, the apparent viscosity increases, but $E_a$ stays the same for the high temperature region (T $\geq$ 20$^\circ$C). The results shown here further suggest that there is a slowing down of water dynamics near the protein as opposed to a strengthening in the H bonds present, and this is attributed a decrease in configurational entropy. Otherwise, we should see an increase in the slope of Fig. \ref{fig:lneta}. This entropy, as mentioned previously, is included in the pre-factor $\eta_0$. This result is consistent with previous studies that conduct similar Brillouin studies where the activation energy was determined for a number of different concentrations of diluted water–tert-butyl alcohol solutions \cite{lupi2011}. Like the work presented in this manuscript, Ref. \cite{lupi2011} also showed that the activation energy E$_a$ for all dilutions had the same value. 

The intermediate temperature region (-2.5$^\circ$C $\leq$ T $\leq$ 25$^\circ$C) displays a non-Arrhenius behaviour as is apparent in Fig. \ref{fig:lneta}. Previously it was stated that the reasoning behind this could be due to the onset of cooperative motions at molecular level associated with a power law, or a Vogel–Fulcher-Taumann global dependence of viscosity \cite{lupi2011}. Previous results obtained for snail mucus suggested that a factor that could be influencing the behaviour seen here could be attributed to the difference in density fluctuation relaxation times and structural relaxation times \cite{lupi2011,comez2012}. 

Additionally, in the region below the freezing point, we can only say that the likely rapid increase in $E_a$ is due the coexistence of ice and mucus, which has previously been reported \cite{hanlon2023temperature}. 

% Thus, with the current data we have in this region and with the results for snail mucus that are already available, we cannot suggest any further hypothesis about the molecular behaviour in this region. 
 
\begin{table}[!t]
\caption{Best-fit parameters for fit of function $\displaystyle \ln (\eta) = \ln \eta_0 +E_a/k_B T$ to experimentally determined apparent viscosity.}
\setlength{\extrarowheight}{3pt}
\begin{ruledtabular}
\begin{tabular}{c c c c c c}
Temperature & Dilution & $\ln \eta_0$ & $E_a$ & R$^2$\\
Range & \# & (cP) & kJ/mol &  \\
 \hline
& 1 & -5.2& 13.1 $\pm$ 0.2 & 0.992\\
25.0$^\circ$C $\leq T \leq$ 52.0$^\circ$C & 2 & -5.4 & 13.3 $\pm$ 0.3 & 0.993\\
& 3 &  -5.1 & 13.3  $\pm$ 0.3& 0.994\\ 
& 4 & -5.2 & 13.3 $\pm$ 0.2 & 0.989\\
    \end{tabular}
    \label{tab:ViscoFit}
    \end{ruledtabular}
\end{table}

\subsubsection{Hypersound Attenuation}
% \begin{figure}[t]
% \centering
% \includegraphics[width=1\linewidth]{SoundAbs_Dilutions.png}
% \caption{Plot of $\alpha/f^2$ as a function of temperature for all dilutions.}
% \label{fig:SoundAbs_Dilutions}
% \end{figure}

Figure \ref{fig:Mucus_Dilution_Comp_Visco} presents the frequency-independent hypersound absorption coefficient, obtained through Eq. \ref{eq:absorp}, as a function of temperature for diluted mucus. At temperatures above $T = 0$ $^\circ$C, the absorption coefficient generally decreasing with increasing temperature. Conversely, as the temperature decreases, the absorption coefficient tends to increase. However, it is important to note that the situation becomes more complex below $T = 0$ $^\circ$C due to the presence of ice crystallites in the system. The presence of these crystallites introduces scattering effects that are not accounted for in the sound absorption formula. The observed decrease in sound absorption with increasing temperature provides valuable insights into the dynamic behavior of the glycoprotein-water structure. It suggests that temperature has an influence on the acoustic properties of the system, indicating there are structural changes occurring in the system, although these changes are likely to be minimal over the dilution range studied. This implies that the dilution of the system, within the investigated range, is not heavily influenced by the addition of water to the system. Other factors, such as the composition and concentration of glycoproteins, or the presence of other impurities may have a stronger influence of the sound absorption.

% The sound absorption for the diluted mucus is complementary to the sound absorption obtained for pure mucus previously obtained \cite{hanlon2023temperature}. Thus over the dilutions studied, the increase in water content does not appear to have a significant influence on the damping of sound waves.

% This is in line with both pure snail mucus and ultra pure water obtained previously \cite{hanlon2023temperature}. 

% Hypersound absorption is the process in which sound waves are absorbed via a number of processes such as thermal relaxation, defects (or impurities), and by viscous dissipation \cite{east1969}. 

\subsubsection{Storage \& Loss Modulus}
\begin{figure}[h]
\centering
\includegraphics[width = 1\linewidth]{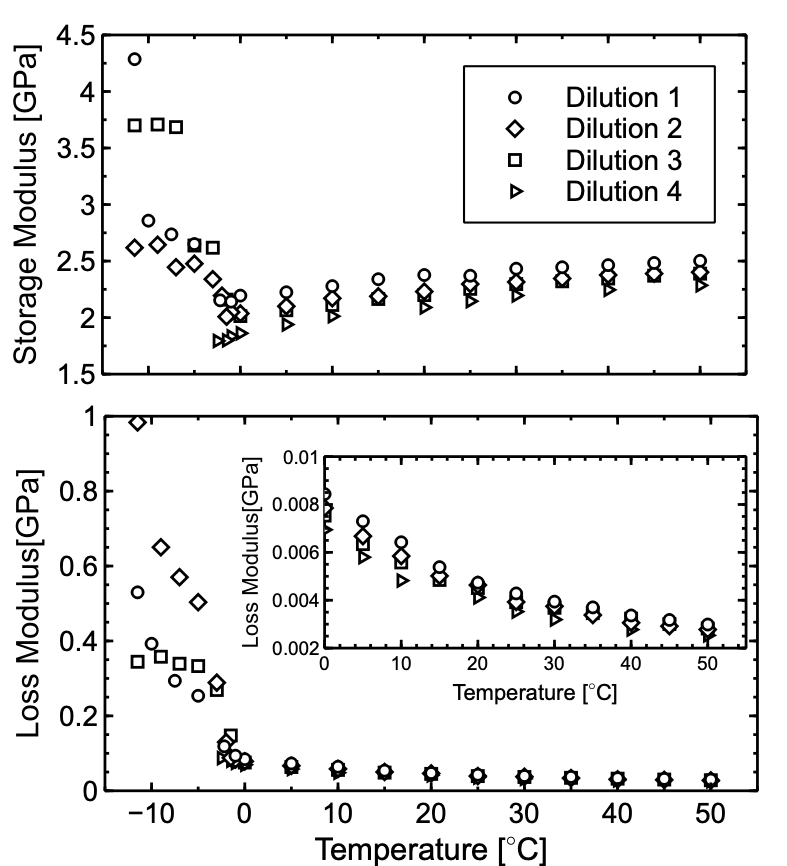}
\caption{Room temperature Brillouin spectra collected on natural snail mucus dehydrating as a function of time. Arrow indicates direction of increasing time.}
\label{fig:ComplexMod_Visco}
\end{figure}

Figure \ref{fig:ComplexMod_Visco} shows the storage modulus calculated using the first term in Eq. \ref{eq:longmod} for all dilutions as a function of temperature. In general, for $T \geq 0$ $^\circ$C the storage modulus shows a gradual increase in value for all dilutions with increasing temperature. Below $T = 0$ $^\circ$C, the storage modulus for dilutions 1-3 behave similarly, all showing a rapid increase with decreasing creasing temperature. However, the storage modulus for dilution 4 begins to decrease with decreasing temperature, similar to the behaviour previously observed for water \cite{hanlon2023temperature}.

Figure \ref{fig:ComplexMod_Visco} also shows the loss modulus calculated used the second term in in Eq. \ref{eq:longmod} for all dilutions as a function of temperature. The loss modulus exhibits a general decrease in value for $T \geq 0$ $^\circ$C for all dilutions. Furthermore, the loss modulus for dilutions 1-3 all increase with decreasing temperatures for $T \leq 0$ $^\circ$C. Dilution 4 also increases with decreasing temperature, but we have limited data for this dilution below 0 $^\circ$C liquid mucus Brillouin peaks became very weak.

\subsubsection{Freezing Point Depression}
\begin{figure}[t]
\centering
\includegraphics[scale = 0.4]{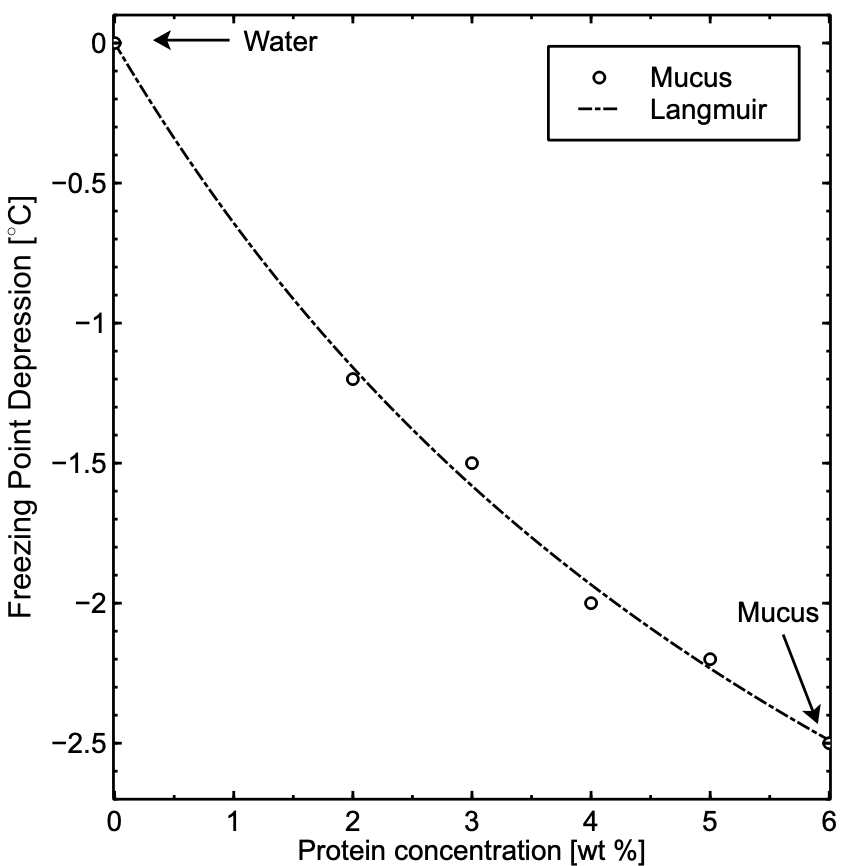}
\caption{Plot of freezing point depression as a function of protein concentration. Dashed curve: Best fit of Eq.~\ref{eq:Langmuirmodel} to the experimental data.}
\label{fig:FreezingPDep}
\end{figure}
\noindent
Figure \ref{fig:FreezingPDep} illustrates the relationship between protein concentration (wt \%) and freezing point depression. The observed behavior aligns with the Langmuirian adsorption model, previously used to explain the adsorption of antifreeze glycoproteins onto ice crystal surfaces \cite{burcham1986kinetic}. This model can be described by the following expression,
\begin{equation}
    \Delta T = \frac{\Delta T_m K c}{1 + Kc}
    \label{eq:Langmuirmodel}
\end{equation}
where, $\Delta T$ represents the freezing point depression, $\Delta T_m$ is the maximum freezing point depression, $K$ denotes the intensity of interaction between glycoprotein and ice, and $c$ represents the concentration. Although the Langmuir model is typically used to describe the adsorption of molecules onto solid surfaces, we find it suitable for our system due to its ability to characterize and consider solute-solvent interactions. Consequently, we fit Equation \ref{eq:Langmuirmodel} to our freezing point depression data, and the resulting fit is displayed in Figure \ref{fig:FreezingPDep}. The fitting yields $\Delta T_m = -5.85^\circ$C and $K = 0.123$ (wt\%$^{-1}$).

% Additionally, we explored alternative models derived from the Langmuir model, namely the Freundlich and Redlich-Peterson models, which demonstrate comparable experimental fits \cite{freundlich1922,kumar2010continuous}. These alternative fits are also presented in Figure \ref{fig:FreezingPDep}. \hl{$\leftarrow$  If you are going to include these other two models, then you need to provide some rationale for doing so - not much, but a sentence or two.  Is there any reason to believe that one of these other models might be more applicable to the mucus system?  If not, remove.}

The trend that we see here in the freezing point depression is similar to previously studied sugar and water solutions and also similar to previous antifreeze glycoprotein studies \cite{uraji1996freezing, ebbi2010}. Although the relationship between the freezing point shown here for snail mucus and the sugar solutions are similar, the concentrations required for sugars to deplete ice growth required much higher sugar concentrations to depress the freezing point by a similar amount to that of snail mucus. In fact, it took nearly five times as much sugar as it did glycoproteins to depress the freezing point to a value equal to that in snail mucus \cite{uraji1996freezing}. The freezing point depression, previously reported for antifreeze glycoproteins showed an overall increase as the concentration increased. As well, the magnitude of the increase was much higher for lower concentrated solutions similar to the snail mucus. This is an indication that these glycoprotein systems influence the system in a similar way. 

\section{Dehydration Experiments}
\subsection{Sample Preparation}
The dehydrated mucus was prepared by initially adding $\sim2$ ml of natural snail mucus to a clear cuvette. The snail mucus was left unsealed at one end and was allowed to dehydrate while room temperature Brillouin spectra was collected. Temperature dependent Brillouin spectroscopy could not be completed due to the current limitations with the temperature-controlled sample chamber.

\subsection{Results: Raw Spectral Signatures}

\begin{figure}[h]
\centering
\includegraphics[width = 1\linewidth]{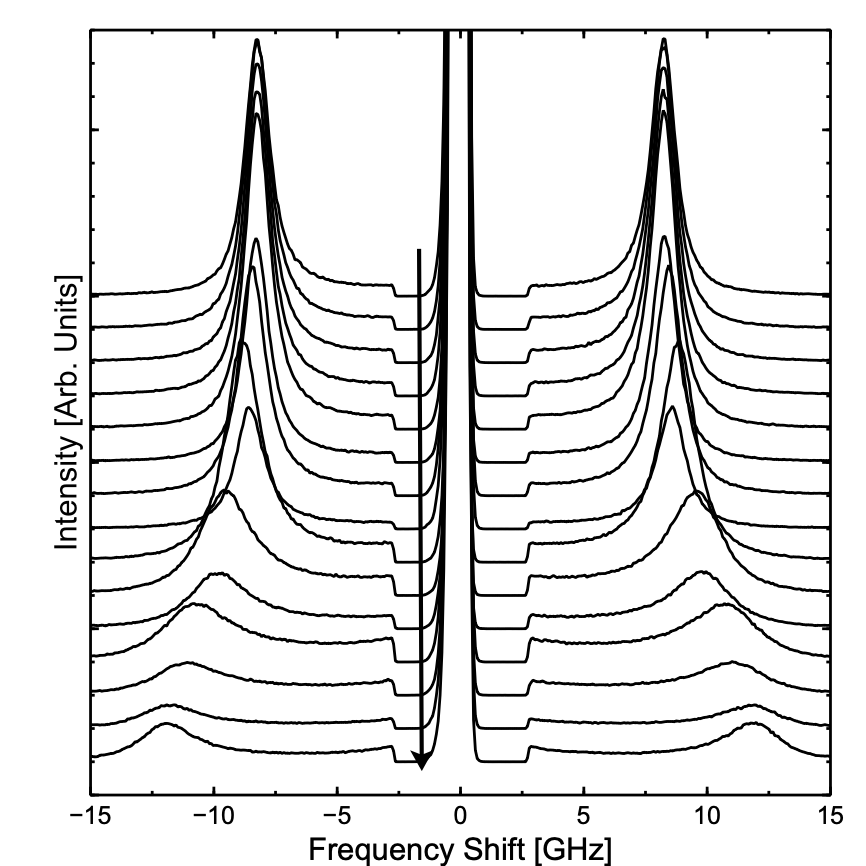}
\caption{Room temperature Brillouin spectra collected on natural snail mucus dehydrating as a function of time. Arrow indicates direction of increasing time.}
\label{fig:BLS_dehydrated}
\end{figure}

Figure \ref{fig:BLS_dehydrated} shows room temperature Brillouin spectra collected on natural snail mucus as a function of time. This series of spectra while although it is a function of time, it is inherently a function of protein concentration as determined by the FTIR data (see Fig \ref{fig:FTIR_Time}). Only a single peak was found in all spectra collected for the dehydrating mucus. However, preliminary dehydration experiments had shown that for a completely dried snail mucus sample, there was the presence of additional peaks. For the purpose of this manuscript this will not be examined here. 

\begin{figure}[t]
\centering
\includegraphics[width=1\linewidth]{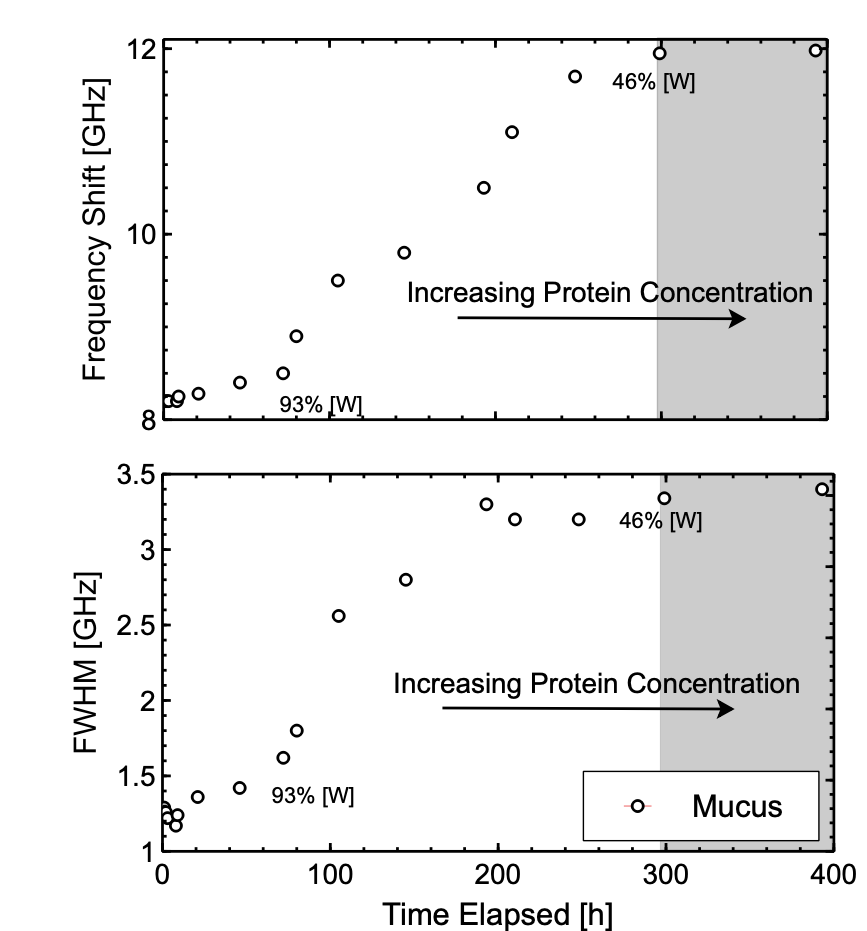}
\caption{
Plot of frequency shift and storage (top plot) modulus, and FWHM and apparent viscosity (bottom plot) as a function of time for the dehydrated mucus sample. Shaded region indicates the transition from a high to low hydrated state.}
\label{fig:Freq_time_dehydrated}
\end{figure}
Figure \ref{fig:Freq_time_dehydrated} shows the frequency shift for the dehydrated mucus as a function of time (concentration). As can be seen from this plot, there are two points where there is a dramatic change in the frequency shift. The first occurring at $\sim$ 75 hours. Before this point, there is an increase in frequency shift from 8.1 to 8.5 GHz, or at a rate of $\sim$ 0.004 GHz/h. After this first point, there is a sudden increase in the frequency shift at a rate of $\sim$ 0.02 GHz/h up to a shift of $\sim$ 12 GHz at 300 hours, as indicated by the dashed line in Fig. \ref{fig:Freq_time_dehydrated}. The second point occurs at a shift of 12 GHz where the frequency then remains constant. The general increase in frequency shift with increasing polymer concentration (time) is expected and consistent with previous Brillouin scattering studies on aqueous polymer solutions \cite{bailey2019brillouin,bedborough1976brillouin, bot1995brillouin,palo2019}.

Figure \ref{fig:Freq_time_dehydrated} shows the FWHM as a function of time for dehydrated mucus. Like the frequency shift data, there exists two points where there is a sudden change in the FWHM of the dehydrated mucus. The first point also occurs at $\sim$ 75 hours, the same as the frequency shift. Above this point, like the frequency shift, there is a linear increase up to the the first transition point. Below the transition, the FWHM increases rapidly. However unlike the frequency shift result, the second transition point occurs at about 200 hours. The FWHM has been previously shown to be heavily influenced by polymer concentration \cite{bailey2019brillouin, van2000structural}. This polymer influence has been proposed previously by other Brillouin studies to the fact that waters mobility is restricted in the presence of polymers hydration shell \cite{lupi2011,comez2012}. 

It has been shown previously in similar Brillouin scattering studies on polymer-water systems that as the concentration in these systems increases, there is a transition from a high-hydration (liquid-like) to a gel like state that occurs \cite{bailey2019brillouin, palo2019}. The shaded region indicated in Fig.~\ref{fig:Freq_time_dehydrated} illustrates the point and region at which we believe this system transitions from a high hydration to a low hydration state. This transition will be further investigated in Section \ref{sec:GeneralDiscussion}. 

\subsection{Results: Viscoelastic Behaviour}
\subsubsection{Sound Velocity}
\begin{figure}[t]
\centering
\includegraphics[width = 1\linewidth]{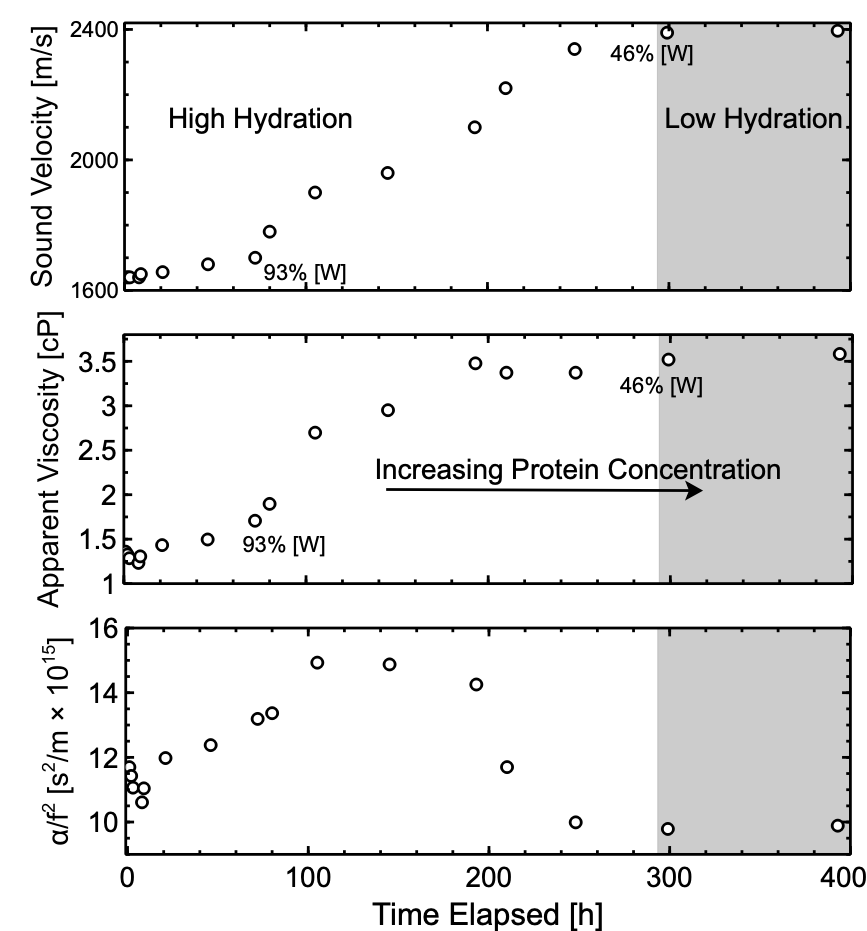}
\caption{Sound velocity, apparent viscosity and sound absorption as a function of time for the dehydrated mucus sample. Shaded region shows the transition from a high to low hydrated state.}
\label{fig:Mucus_Dehydration_CompPlot_Visco}
\end{figure}

Figure \ref{fig:Mucus_Dehydration_CompPlot_Visco} shows the sound velocity as a function of time for the dehydrated mucus. Just like the frequency shift and FWHM results displayed in Fig. \ref{fig:Freq_time_dehydrated}, there exists two transition points where the behaviour of sound velocity changes. The first is at 75 hours, where before this transition point, there is a slight linear increase from $\sim$ 1600m/s to 1700 m/s. After the transition the sound velocity displays a sharp linear increase in value from 1700 m/s to 2400 m/s where the second transition occurs. After the second transition at 300 hours the sound velocity plateaus and maintains a constant velocity of $\sim$ 2400 m/s. This result is an indication that as the mucus dehydrates the structure of this system is becoming more rigid and ``gel-like". If this were not the case, then one would expect the sound velocity to increase only slightly since for liquids, especially those with high water content have sound velocities comparable to that of water \cite{hanlon2023temperature, lupi2011, comez2012}. 

% Additionally, the sound velocity is required for calculation of the storage modulus $M^{\prime}$. It should be noted that in Fig. \ref{fig:Soundvel_dehydrations} the shaded region indicates the low-hydration state discussed previously.

\subsubsection{Apparent Viscosity}
Figure~\ref{fig:Mucus_Dehydration_CompPlot_Visco} shows the natural logarithm of the apparent viscosity for dehydrated snail mucus calculated using Eq.~\ref{eq:longmod}. The apparent viscosity data exhibits two distinct transition points at approximately 75 and 200 hours. Above the first transition, there is a gradual linear increase in apparent viscosity. Subsequently, the apparent viscosity continues to rise linearly until reaching a value of approximately 3.5 cP at the second transition point. Beyond this second transition, the apparent viscosity remains relatively constant.

This data indicates that as the elapsed time increases, the viscosity generally increases. Furthermore, the behavior of the apparent viscosity  differs above and beyond the first transition point, indicating the occurrence of a structural transition. These changes in apparent viscosity provide clear evidence of modifications in the structure of the glycoprotein-water network. The increase in viscosity might be associated with factors such as enhanced intermolecular interactions or strengthening of the cross-linked glycoprotein-water network.

Additionally, the fact that the viscosity remains constant after the second transition point suggests that the structure has reached a stable state and is not easily able to change. 

\subsubsection{Hypersound Attenuation}
Figure~\ref{fig:Mucus_Dehydration_CompPlot_Visco} shows the hypersound absorption coefficient for dehydrated mucus calculated using Eq. \ref{eq:absorp}. The observed behavior of the hypersound absorption coefficient for dehydrated mucus is unlike the previous data shown in this section, the hypersound absorption shows an unusual behaviour. Initially, there is a gradual increase in the damping of sound waves, reaching a maximum absorption point at approximately 100 hours. Following this, there is a gradual decrease leading to a minimum hypersound absorption at around 250 hours, which then remains constant. This unusual behavior of the hypersound absorption highlights the complex nature of mucus dehydration and its impact on sound wave propagation properties. 

This result is rather intriguing and suggests that as the system transitions to a gel-like state, the sound absorption by the glycoprotein-water network decreases. One possible reason that may account for the sound absorption in this is the increased structure associated with the gel state. In a liquid state, the molecules are relatively more mobile and can easily vibrate in response to sound waves. This molecular motion leads to effective sound energy dissipation and absorption. However, as the system transitions to a gel-like state, the increased molecular interactions and structural rigidity can restrict the movement of molecules. This restricted mobility reduces the ability of the gel state to effectively absorb sound energy.

% This result shows that as the concentration increases, the hypersound absorption reaches a maximum at ~11\% protein (89\% water) concentration. This is also the approximate concentration when there is a rapid increase in the rate at which the frequency shift begins to increase. We know that at $\sim$ 54\% protein (46\% water) concentration the system transitions to a gel-like state so perhaps the minimum protein concentration to begin the phase transition is at $\sim$ 11\%. This would coincide with the data displayed throughout this work since this is the approximate concentration at which we see obvious changes in the frequency shift, FWHM, apparent viscosity, and now the hypersound absorption. Furthermore, looking at the region $\geq$ 11\% protein concentration, the data displays an exponential like behaviour as previously seen in the temperature dependent hypersound absorption data for the diluted mucus (see Fig \ref{fig:SoundAbs_Dilutions}).

\subsubsection{Storage \& Loss Modulus}
\begin{figure}[t]
\centering
\includegraphics[width = 1\linewidth]{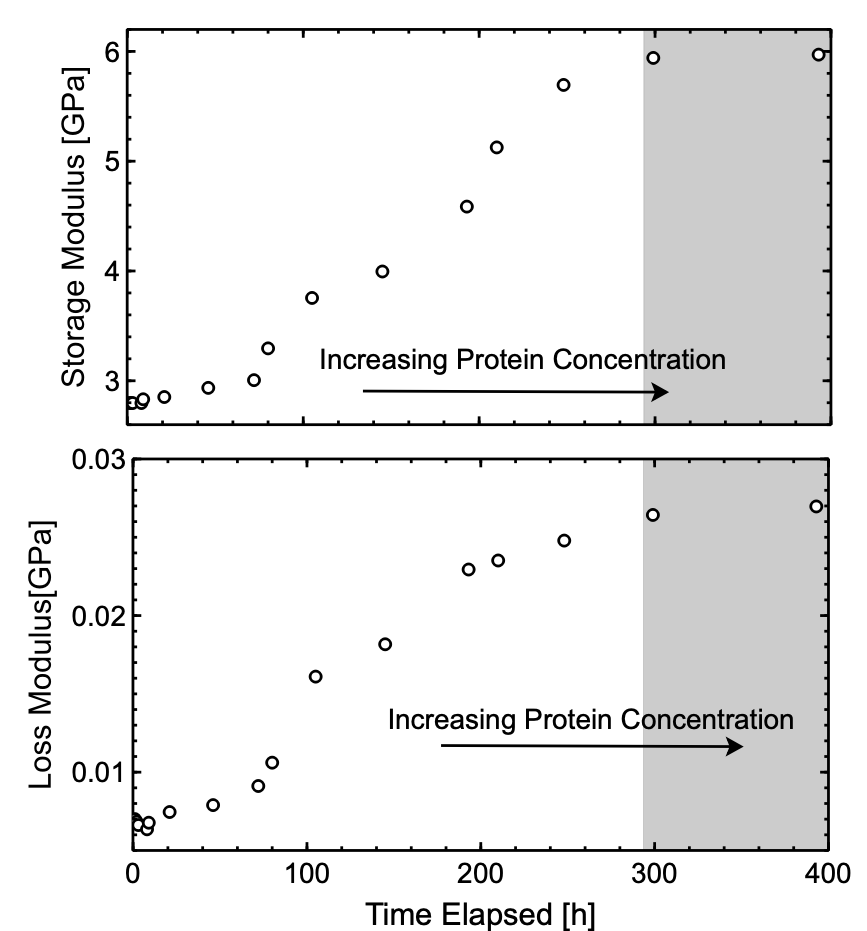}
\caption{Storage and Loss modulus of dehydrating snail mucus as a function of temperature.}
\label{fig:StorageLossModulus_dehydration}
\end{figure}
Figure \ref{fig:StorageLossModulus_dehydration} shows the storage modulus as a function of temperature calculated using the first term in Eq. \ref{eq:longmod}. The storage modulus shows two transition points just like the other quantities detailed in this work occurring at 75 and 300 hours. Before the first transition, the storage modulus increases linearly from $\sim$ 2.9 GPa at 0 hours to about 3.0 GPa at 75 hours. After the first transition, the storage modulus increases rapidly, up to about 6 GPa where it approaches the second transition. After the second transition, the storage modulus remains constant for the entirety of this experiment. The storage modulus as we know, is a measure of how much energy is stored elastically in the system. From this, Fig. \ref{fig:StorageLossModulus_dehydration} tells us that as the protein concentration increases, there is a general increase in the energy stored elastically in the mucus, where it eventually reaches a maximum value of 6 GPa at 300 hours. 

Similarly, Fig.~\ref{fig:StorageLossModulus_dehydration} also shows  the loss modulus calculated using the second term in Eqn. \ref{eq:longmod} as a function of temperature. Like the storage modulus, the loss moduli shows a very similar behaviour. That is, there are two transition present, both also at 75 hours and 300 hours. Before the first transition, there is a linear increase from 0.005 to 0.01 GPa in the loss modulus. After the first transition, there is non-linear increase up to a value of 0.03 GPa at 300 hours. Below the second transition (at 300 hours) the loss modulus appears to slightly increase. The loss modulus is a measure of how much energy is lost through heat in a system, what the data presented here shows is that as the protein concentration increases, the amount of energy lost through heat also increases. 

The result from a previous Brillouin scattering study on the protein concentration of a collagen gel shows a similar plot to Fig. \ref{fig:StorageLossModulus_dehydration}. However, as suggested by this study, the loss modulus should exhibit an increase in value once it transitions to a gel like state which is present in the data for loss modulus, albeit slight.

\section{General Discussion}
\label{sec:GeneralDiscussion}
% \hl{I have not yet read this section, but I noticed that it has no figures.  I envision this section as containing at least a schematic figure or two presenting a big picture view of what the combined results of your experiments tell you about the behaviour/happenings in the mucus system. For example, plots containing ALL (diluted and dehydrated) data (shift, FWHM, velocity, viscosity, ...) at room T vs water/protein concentration would be extremely insightful. Think about this and try to incorporate to aid discussion in text.  See my earlier note below (in source file). :-)}

\subsection{Structural Transition: Influence of Glycoproteins}

%  \begin{figure}[!h]
% \centering
% \includegraphics[width = 1\linewidth]{Freq_FWHM_Concentration.png}
% \caption{Room temperature Brillouin frequency shift, and FWHM as a function of protein concentration.}
% \label{fig:Freq_FWHM_Concentration}
% \end{figure}

%  \begin{figure}[!h]
% \centering
% \includegraphics[width = 1\linewidth]{Mucus_Dilute_Conc_Comp.png}
% \caption{Room temperature Brillouin frequency shift, and FWHM as a function of protein concentration.}
% \label{fig:Visco_Concentration}
% \end{figure}

\begin{figure*}[!t]
\centering
\includegraphics[scale = 0.55]{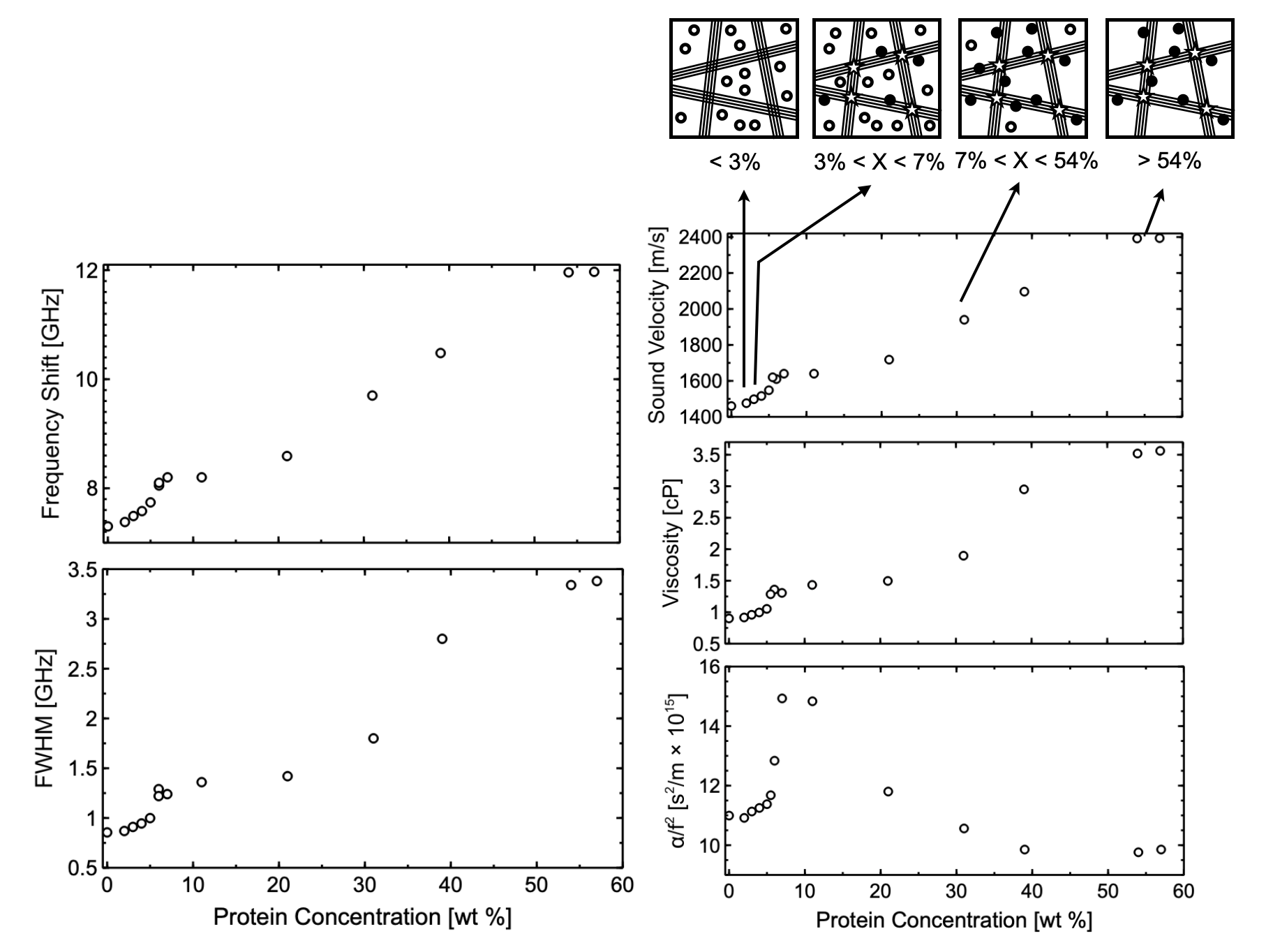}
\caption{Plots of room-temperature Brillouin peak frequency shift and FWHM (left panel) and derived viscoelastic properties (right panel) as a function of protein concentration in snail mucus solutions. Additionally, a simple schematic is provided to illustrate the observed structural transitions. These transitions, along with the concentration at which they occur is depicted in the right panel. The symbols in this figure correspond to $\circ$ - free water, $\bullet$ - bound water, $\medwhitestar$ - cross-links and lines represent glycoproteins.}
\label{fig:Transition_Schematic}
\end{figure*}

This study on natural snail mucus has revealed the presence of at least three structural transitions, each characterized by a discontinuous change observed in both raw Brillouin scattering data and calculated viscoelastic properties. These transitions mark significant shifts in the physical properties of the mucus, indicating changes in the molecular arrangement, intermolecular forces or conformational dynamics in the glycoprotein-water network. 

Figure \ref{fig:Transition_Schematic} shows the frequency shift and FWHM of both diluted obtained from room temperature Brillouin spectroscopy of both diluted and dehydrated snail mucus, plotted as a function of protein concentration. As mentioned above, we observe three distinct transitions in the raw spectral data, along with calculated quantities throughout this manuscript. The first transition occurs at approximately 3\% protein concentration. Both the frequency shift and FWHM show a sharp increase in value at this point. Despite only having one data point for diluted mucus (dilution 4) above this point, the frequency shift and FWHM is significantly different than dilutions 1 - 3. Additionally, if we consider the temperature dependence of frequency shift and FWHM shown previously (see Fig \ref{fig:Freq_time_dehydrated}) we can see that data for dilution 4 resembles that of pure water obtained previously by Brillouin spectroscopy \cite{hanlon2023temperature}. The dilution data presented throughout this work suggests that this first transition is likely attributed to the proteins becoming first cross-linked in the system. Further support for this is through the calculated viscoelastic properties, speed of sound, apparent viscosity and sound absorption as shown in the right panel of Fig. \ref{fig:Transition_Schematic}. Furthermore, the temperature dependence of the viscoelastic properties for each dilution previously shown (see Figures \ref{fig:Mucus_Dilution_Comp_Visco} - \ref{fig:ComplexMod_Visco}) all increase in value as the protein concentration increases, a particularly strong increase is observed between 2 and 3\% protein concentration. Perhaps, the most obvious increase in the viscoelasticity is observed in Fig. \ref{fig:lneta}, where there is a clear difference between the natural logarithm of apparent viscosity for dilution 3 and 4. Recall, that this change was associated with the entropic pre-factor $\eta_0$, which is subject to conformational changes \cite{lupi2011,comez2012}. It has been stated before that cross-linked protein-water systems increase the elastic and acoustic properties of the system \cite{hanlon2023temperature,bailey2019brillouin,denn1980,denn1984} which would also support our idea proposed here.

The second transition we observe occurs at a protein concentration of approximately 7\% and has been attributed to the depletion of any free water in the snail mucus. Bound and free water exist in polymer gel networks naturally and together make up the entire water content in such systems \cite{peppas1977development,peppas1993preparation,hoffman2012hydrogels}. Bound water refers to the water molecules that are tightly associated or ``bound" to the protein-water network, while free water, on the other hand, represents the water that is not bound to the network and is ``free" to move in the system. The greater amount of bound water in a system can increase the cross-linking density which in turn causes the molecular structure more rigid \cite{hoffman2012hydrogels}. The observed behavior in both the raw Brillouin data and calculated viscoelastic quantities supports the notion that this transition is caused by the initial depletion of free water in the mucus. First considering, Fig. \ref{fig:Transition_Schematic}, we can see a general increase in both the frequency shift and FWHM as the protein concentration increases after a protein concentration of 7\% up to $\sim$ 50\%. The increase in both frequency shift and FWHM with protein concentration in this region is a clear indication of increased rigidity. Additionally, the viscoelastic properties presented in Fig. \ref{fig:Transition_Schematic} show an increase in the sound velocity and viscosity with increasing concentration, while $\alpha/f^2$ decreases with increasing concentration. The increase in both sound velocity and viscosity observed in this region signifies a significant shift in the molecular structure, indicating a transition towards increased rigidity. The rise in sound velocity, is a measure of how quickly sound propagates through a medium, highlighting the increased stiffness of the glycoprotein-water network. Simultaneously, the increased apparent viscosity, a measure of a fluids resistance to flow, further supports this transition by indicating a greater resistance to molecular rearrangements. Moreover, the decrease in the sound absorption in this region provides more justification for this claim. In pure liquids, molecules have higher mobility, allowing them to vibrate and dissipate sound energy effectively. The random arrangement of molecules in a liquid state provides more opportunities for sound waves to interact with the medium and be absorbed. Thus, as the protein concentration in the system increases, and the internal molecular structure becomes more rigid, then the molecular mobility will be restricted leading to a lower sound absorption.

Lastly, the third transition observed in this work occurs at a protein concentration of 54\% and has been deemed due to the transition to a gel state and it is suggested that some of the remaining bound water be evaporated at this state. As illustrated by the raw spectral data in Fig. \ref{fig:Transition_Schematic}, there is a clear and distinct transition occurring at a protein concentration of 54\%, where both the frequency shift and FWHM go from increasing to a constant value. Due to the relatively high frequency shift and FWHM in this region, along with the fact that there is still a considerable amount of water left in the system, as determined by the ATR data, this transition was therefore attributed to a gel-like state. Further support for this is based on the calculated viscoelastic properties shown in Fig. \ref{fig:Transition_Schematic}. Both, the sound velocity, viscosity and sound absorption all show their respective values remain constant for this high protein concentration region. One would expect that for a gel state, the structure be more rigid mainly for two reasons. Firstly, a decrease in water content typically results in increased rigidity of the gel-like structure. Secondly, the reduction in free water, with only bound water being incorporated into the glycoprotein network, can contribute to an increase in molecular arrangement structural and therefore greater rigidity. In the gel state, as the free water content decreases, the molecular interactions among glycoproteins and bound water become more pronounced. This increased interaction contributes to a more structured and rigid gel network. Moreover, the presence of bound water molecules in the gel network restricts the movement of proteins and other components. Previous Brillouin scattering studies on gelatin gels as a function of protein concentration show behaviour nearly identical to what is shown here \cite{bailey2019brillouin}. In that Brillouin study, there is a plateau at high concentration values, like that shown here. Likewise, the FWHM shows a brief plateau at concentrations near 50\% protein, similar to our results. 

% Figure \ref{fig:Transition_Schematic} presents a visual representation of the three distinct transitions observed in this study. In Fig. \ref{fig:Transition_Schematic}b, it portrays the concentration at which free water depletion commences, resulting in an increased structure of glycoprotein-bound water. Lastly, in Fig. \ref{fig:Transition_Schematic}c, the final transition observed at a protein concentration of 54\% is depicted. At this stage, all free water has evaporated, leaving behind a gel state composed of glycoprotein and water bound to the cross-linked network.

\section{Conclusion}

Brillouin light scattering spectroscopy was employed to investigate the viscoelastic properties and phase behavior of diluted and dehydrated snail mucus. The temperature range of -11$^\circ$C $\leq T \leq$ 52 $^\circ$C was explored for all dilutions of the snail mucus, while the dehydrated mucus was examined with respect to time and corresponding protein concentration. Intriguingly, deviations in both temperature and concentration dependencies of spectral peak parameters and derived viscoelastic properties validate the previously identified liquid-to-solid phase transition at -2.5$^\circ$C, while additionally revealing a correlation between concentration and this transition temperature T$_{pt}$. Collectively, both the dilution and dehydration data suggest three distinct structural transitions occurring at protein concentrations of 3\%, 7\% and 54\%. These transitions were attributed to the initiation of cross-linking in the glycoprotein water network, depletion of free water, and the transition to a gel state. These results support previous Brillouin studied on gels and provide further evidence for the intricate role of glycoproteins within these systems \cite{bailey2019brillouin}. Notably, this research offers fresh insights into the significance of hydration water in protein-water systems of biological relevance. Furthermore, it underscores the effectiveness of Brillouin spectroscopy in studying this unique class of natural materials, expanding the range of systems to which this technique has proven to be highly valuable.

\bibliography{bibliography}

\end{document}